\documentclass{amsproc}

\newtheorem{theorem}{Theorem}[section]
\newtheorem{lemma}[theorem]{Lemma}

\theoremstyle{definition}

\theoremstyle{remark}

\newcommand{\cc}{\mathcal{C}}
\newcommand{\m}{\mathcal{M}}
\newcommand{\D}{\mathcal{D}}

\newcommand{\dbar}{\overline{\partial}}

\numberwithin{equation}{section}

\begin{document}

\title{Solution of the initial value problem for the focusing Davey-Stewartson II system}

\author{ E. Lakshtanov}
\address{Department of Mathematics, Aveiro University, Aveiro 3810, Portugal.}
\email{lakshtanov@ua.pt}
\thanks{The first author was supported by Portuguese funds through the CIDMA - Center for Research and Development in Mathematics and Applications and the Portuguese Foundation for Science and Technology (``FCT--Fund\c{c}\~{a}o para a Ci\^{e}ncia e a Tecnologia''), within project UID/MAT/0416/2013.}

\author{B. Vainberg}
\address{Department
of Mathematics and Statistics, University of North Carolina,
Charlotte, NC 28223, USA.}
\email{brvainbe@uncc.edu}
\thanks{The second author was supported  by the NSF grant DMS-1410547}

\subjclass{Primary 54C40, 14E20; Secondary 46E25, 20C20}
\date{October 8, 2017.}

\dedicatory{Dedicated to a remarkable person
                           and mathematician,  Prof. S. Krein.}

\keywords{Davey-Stewartson, $\dbar$-method, Dirac equation, exceptional point, scattering data, inverse problem}


\begin{abstract}
We consider a focusing Davey-Stewartson system and construct the solution of the Cauchy problem in the presence of exceptional points (and/or curves).
\end{abstract}

\maketitle

\section{Introduction}
Let $q_0(z), ~ z\!=\!x\!+\!iy, ~ (x,y)\! \!\in \!\mathbb R^2$, be a compactly sufficiently smooth function. Consider the DSII system of equations for unknown functions $q=q(z,t),~\phi=\phi(z,t),  ~x,y \in \mathbb R^2, ~ t\geq 0:$
\begin{eqnarray} \nonumber
q_t=2iq_{xy} -4i \phi q, ~\\ \nonumber
\phi_{xx}+\phi_{yy} = \pm 4 |q|^2_{xy}, \\
q(z,0)=q_0(z).\label{dsii}
\end{eqnarray}
   The sign plus in (\ref{dsii}) corresponds to the defocusing case of DSII and sign minus corresponds to the focusing case. Even though the defocusing case is well studied, we will consider both models together to stress the universality of our approach.

The Davey-Stewartson system of equations models the shallow-water limit of the evolution of weakly nonlinear water
waves that travel predominantly in one direction, but in which the wave amplitude
is modulated slowly in two horizontal directions \cite{ds}.  The shallow-water limit means that $kh \rightarrow 0$, $a \ll kh^2$,  where $h$ denotes the depth of the bed, $k$ is a wave number of a wavy surface and $a$ is a characteristic amplitude of the disturbance.

Independently, Ablowitz and Haberman \cite{ah}, Morris \cite{morris}, and Cornille \cite{corn} have
derived  (\ref{dsii}) while looking for completely integrable systems generalizing the nonlinear Schr\"{o}dinger equation to two spatial dimensions.

In \cite{saut2}, \cite{saut1}, one can find results on global uniqueness and existence of solutions of  (\ref{dsii}) in the defocusing case and for small initial data in the focusing case, as well as a justification of local in time well-posedness for arbitrary data in both cases.

Constructions of solutions of initial value problem (\ref{dsii}) via the IST (inverse scatteing transform) in the defocusing case or when initial data are small enough are given in \cite{bc1}, \cite{bc2}, \cite{sung1}-\cite{sung3}, \cite{perry}. These works used the $\overline \partial$-method for the Dirac inverse scattering problem.
The classical $\overline \partial$-method fails when exceptional points are present. The latter are defined as the values of the spectral parameter {k} for which the homogeneous direct scattering problem has a non-trivial solutions. In \cite{lnv}, \cite{lbcond}, we generalized
the $\overline \partial$-method for the Schr\"{o}dinger and Dirac equations to the case when exceptional points exist.
A prototype of this generalization was considered in section 8 of \cite{RN2}.
In the current paper, we apply the results of \cite{lbcond} to solve the Cauchy problem for the focusing DSII.

We will work with equation (\ref{dsii}) rewritten in the following form (see, for example, \cite{fokasp}):
\begin{eqnarray}\nonumber
q_t=2iq_{xy}  \pm 4q(\overline{\varphi} -\varphi), ~\\ \nonumber
\partial \varphi =\dbar |q|^2, \\
q(z,0)=q_0(z).\label{dsiiB}
\end{eqnarray}
In order to obtain (\ref{dsii}) from (\ref{dsiiB}), one needs to introduce $\phi=i\pm( \overline{\varphi}-\varphi),$
apply the Laplacian to $\phi$, and use that $\partial \dbar= \dbar \partial$.

\section{The main result}\label{sectmr}
Denote
\begin{eqnarray}\label{char1}
Q^0(z)=\left ( \begin{array}{cc}0 &q_0(z) \\
\pm q_0(z) &  0\end{array} \right ),   \quad  z \in \mathbb C.
\end{eqnarray}
Consider the Dirac equation for the $2\times 2$ matrix $\psi(\cdot,k), ~ k \in \mathbb C:$
\begin{equation}\label{fir}
\frac{\partial \psi}{\partial \overline{z}} = Q^0 \overline{\psi}, \quad \psi(z,k) e^{-i\overline{k}z/2} \rightarrow I, \quad z \rightarrow \infty.
\end{equation}
The corresponding Lippmann-Schwinger equation has the following form
\begin{equation}\label{19JanA}
\psi(z,k)=e^{i\overline{k}z/2}I+ \int_{z \in \mathbb R^2} G(z-z',k) Q^0(z') \overline{\psi}(z',k) d{z'}, \quad G(z,k)=\frac{1}{\pi}\frac{e^{i\overline{k}z/2 }}{z}.
\end{equation}
  Solutions $\psi$ of (\ref{19JanA}) are called the {\it scattering solutions}. Let $q_0(z)\in L^p_{\rm{comp}}(\mathbb R^2), ~p>2$. Here and below we use the same notation for functional spaces, irrespectively of whether those are the spaces of matrix-valued or scalar-valued functions.
 After the substitution
\begin{equation}\label{subs}
\mu(z,k) = \psi(z,k)e^{-i\overline{k}z/2},
\end{equation}
equation  (\ref{19JanA}) takes the form
\begin{equation}\label{lsls}
\mu(z,k) = I+ \int_{z \in \mathbb R^2} \frac{e^{-{\rm Re}(i\overline{k}z')}}{\pi(z-z')}Q^0(z') \overline{\mu}(z',k) d{z'},
\end{equation}
where the integral operator is compact in $L^q(\mathbb R^2), ~ q>\frac{2p}{p-2},$ see \cite[Th.A. iii]{bu}.
The values of $k$ such that the homogeneous equation (\ref{lsls}) has a non-trivial solution are called the {\it exceptional points}.  The set of exceptional points  will be denoted by $\mathcal E$. Note that the operators in equations (\ref{19JanA}), (\ref{lsls}) are not analytic in $k$, and $\mathcal E$ may contain one-dimensional components. Thus the scattering solution may not exist if $k\in \mathcal E$.
There are no exceptional points when $|k|$ is large enough (e.g., \cite[Lemma 2.8]{sung1}, \cite[Lemma C]{bu}). Let us choose $A\gg1$ and $k_0 \in \mathbb C$  such that all the exceptional points are contained in the disk
\begin{equation}\label{set28A}
D=\{  k \in \mathbb C: 0 \leq |k| < A\},
\end{equation}
and that $k_0 $ belongs to the same disc $ D$ and is not exceptional.

The generalized scattering data are defined by the following integral
\begin{equation}\label{27dec2}
h_0(\varsigma,k)  =  \frac{1}{(2\pi)^2} \int_{R^2}  e^{-i\overline{\varsigma}z/2}Q^0(z)\overline{\psi}(z,k)dxdy, ~ \varsigma \in \mathbb C,~ k \in \mathbb C\backslash \mathcal E.
\end{equation}
In fact, from the Green formula it follows that $h_0$ can be determined without using the potential $Q^0$ or the solution $\psi$ of the Dirac equation (\ref{fir}) if the Dirichlet data at  $\partial\Omega$ are known for the solution of (\ref{fir}) in a bounded region $\Omega$ containing the support of $Q^0$.

The inverse scattering problem of reconstructing the potential $Q^0$ via the given $h_0$ plays a crucial point in the present paper. This inverse problem has been solved in \cite{bc2, sung1} for symmetric or small antisymmetric $Q^0$ using the $\dbar$-method. With larger potentials,  exceptional points appear, and the $\dbar$-method must be replaced \cite{lnv, lbcond} by a combination of $\dbar$ and the Riemann-Hilbert methods (which we called the global Riemann-Hilbert problem). For the
reconstruction procedure, we consider the space
\[
\mathcal H^s = \left \{ u \in L^s(\mathbb R^2)\bigcap C(D) \right   \}, \quad s>2,
\]
(recall that we use the same notation for matrices if their entries belong to $\mathcal H^s$) and the operator
\begin{eqnarray}\label{0712GAA}
T_{z} \phi (k)=
\frac{1}{\pi} \int_{\mathbb C\backslash D}  e^{i(\overline{\varsigma}z+\overline{z}{\varsigma})/2} \overline{\phi}(\varsigma) \Pi^o h_0(\varsigma,\varsigma)  \frac{d\varsigma_R d\varsigma_I}{\varsigma-k}+\\ \nonumber
\frac{1}{2\pi i} \int_{\partial D} \frac{d\varsigma}{\varsigma - k} \int_{\partial D}
[e^{\frac{i}{2}(\!\varsigma\overline{z}+\overline{\varsigma'}z\!)}\overline{\phi^-(\varsigma')} \Pi^o  + e^{i(\!\varsigma-{\varsigma'}\!)\frac{\overline{z}}{2}}\phi^-(\varsigma')\Pi^d \bold C] \left [{\rm Ln}\frac{\overline{\varsigma'}-\overline{\varsigma} }{\overline{\varsigma'}-\overline{k_0} }h_0(\varsigma',\varsigma) d\overline{\varsigma'} \right ],
\end{eqnarray}
where $z \in \mathbb C, ~ \phi\in \mathcal H^s$,
$\phi^-$ is the boundary trace of $\phi$ from the interior of $D$,
$\bold C$ is the operator of complex conjugation, $\Pi^oM$ is the off-diagonal part of a matrix $M$, $\Pi^dM$ is the diagonal part. The logarithmic function here is well defined, see the Remark
after Lemma 3.4 in \cite{lbcond}.

It turns out that, after the substitution $w=v-I\in\mathcal H^{s}, ~{s}>2,$ the equation
\[
(I+T_z)v=I
\]
becomes Fredholm in $\mathcal H^{s}$, and the potential $q_0$ can be expressed explicitly in terms of $v$ (see \cite{lnv, lbcond}).

In order to solve the DSII problem  (\ref{dsiiB}), we apply this reconstruction procedure to specially chosen scattering data. We start with the generalized scattering data defined by $q_0$ and extend them in time as follows:
\begin{equation}\label{2506B}
h(\varsigma',\varsigma,t):=e^{-t(\varsigma^2-\overline{\varsigma'}^2)/2}\Pi^o h_0(\varsigma',\varsigma) +e^{-t(\overline{\varsigma}^2-\overline{\varsigma'}^2)/2}\Pi^d h_0(\varsigma',\varsigma),~\varsigma'\in\mathbb C, ~ \varsigma \in \mathbb C\backslash \mathcal E, ~ t \geq 0.
\end{equation}
For $t \geq 0$, we define the operator
\begin{eqnarray}\label{0712G}
T_{z,t} \phi (k)=
\frac{1}{\pi} \int_{\mathbb C\backslash D}  e^{\frac{i}{2}(\!\overline{\varsigma}z+\overline{z}{\varsigma}\!)} \overline{\phi}(\varsigma) \Pi^o h(\varsigma,\varsigma,t)  \frac{d\varsigma_R d\varsigma_I}{\varsigma-k}+\\ \nonumber
\frac{1}{2\pi i} \!\int_{\partial D} \!\frac{d\varsigma}{\varsigma - k} \!\int_{\partial D}\!\!
[e^{\frac{i}{2}(\!\varsigma\overline{z}+\overline{\varsigma'}z\!)}\overline{\phi^-(\varsigma')} \Pi^o \! +\! e^{i/2(\varsigma-{\varsigma'})\overline{z}}\phi^-(\varsigma')\Pi^d \bold C] \!\!\left [{\rm Ln}\frac{\overline{\varsigma'}-\overline{\varsigma} }{\overline{\varsigma'}-\overline{k_0} }h(\varsigma',\varsigma,t) d\overline{\varsigma'} \right ].
\end{eqnarray}

Assumptions on $q_0$ (that will be stated later) imply that the equation
\begin{equation}\label{1603B}
 (I+T_{z,t})v_{z,t} = I
\end{equation}
remains Fredholm in the space $\mathcal H^{s}, ~ s>2,$ after the substitution $w_{z,t}=v_{z,t}-I$. Let $\Omega$ be an arbitrary region in the half space $\{(z,t):z\in \mathbb C,t\geq 0\}$, where the kernel of $I+T_{z,t}:\mathcal H^{s}\to \mathcal H^{s}, ~s>2,$ is trivial. We will show that the vector $(q(z,t),\varphi(z,t))$ defined by
\begin{eqnarray}\label{1112A}
\\ \nonumber
\left (  \!\!\! \begin{array}{cc} \varphi(z,t) & q(z,t) \\ \pm q(z,t) & \varphi(z,t) \end{array} \!\! \! \right )
\!:=\!\frac{-i}{2}(\Pi^o + \dbar \Pi^d) \!\left (
 \frac{1}{\pi} \int_{\mathbb C\backslash D} \!\! e^{i(\overline{\varsigma}z+\overline{z}{\varsigma})/2} \overline{v_{z,t}}(\varsigma) \Pi^oh_0(\varsigma,\varsigma,t)  {d\varsigma_R d\varsigma_I} + \right .\\ \left .
\frac{ 1}{2\pi i} \int_{\partial D}\! {d\varsigma} \!\int_{\partial D}
\![e^{\frac{i}{2}(\!\varsigma\overline{z}+\overline{\varsigma'}z\!)}\overline{v_{z,t}^-(\varsigma')} \Pi^o  \!- e^{\frac{i}{2}(\!\varsigma-{\varsigma'})\overline{z}}v_{z,t}^-(\varsigma'\!)\Pi^d \bold C] \left [{\rm Ln}\frac{\overline{\varsigma'}-\overline{\varsigma} }{\overline{\varsigma'}-\overline{k_0} }h_0(\varsigma',\varsigma,t) d\varsigma' \right ] \!\right ),\nonumber
\end{eqnarray}
where $v_{z,t}$ is the solution of (\ref{1603B}), solves the DSII equation in $\Omega$.


We need a couple of definitions before we state the main result.

We'll use the word {\it generic} when referring to elements that belong to an open, dense subset $V$ of a topological space $S$.
Let $ W^{n,\infty}_{\rm{comp}}(\mathbb R^2)$ be the space of compactly supported functions with bounded derivatives of orders $j\leq n$ and with the support in a bounded region $\mathcal O\subset R^2$.

We will say that a set $\omega$ of points $(z,t)$ in $ \mathbb R^3_+=\mathbb R^3\bigcap\{t\geq 0\}$ is half-open if $ \omega $ contains points where $t=0$ and, for each point $(z_0,0)\in\omega$, there is a ball $B_0$ centered at this point such that $B_0\bigcap\{t\geq 0\}\subset\omega$.
\begin{theorem}\label{mthm} Let $q_0(z) \in W^{6,\infty}_{\rm{comp}}(\mathbb R^2)$.   Then, for each  $s>2$, the following statements are valid.
\begin{itemize}
\item
The operator $T_{z,t}$ is compact in $\mathcal H^{ s}$ for all $z \in \mathbb C, ~t \geq 0,$ and depends continuously on $z$ and $t \geq 0$.
The same property holds for its first derivative in time and all the spacial derivatives  (in $\Re z,~\Im z$) up to the third order, where the derivatives are defined in the norm convergence. When (\ref{0712G}) is differentiated, the derivatives can be applied to the integrands. The function $T_{z,t} I$ belongs to $\mathcal H^{ s}$ for all $t \geq 0$.

  \item Let the kernel of $I+T_{z,t}$ in the space $\mathcal H^{s}$ be trivial for $(z,t)$ in an open or half open set $\omega\subset \mathbb R^3_+= \mathbb R^3\bigcap\{t\geq 0\}$. Let $v_{z,t}=w_{z,t}+I$, where $w_{z,t}\in \mathcal H^{s}$ is the solution of the equation
      \begin{equation}\label{inteq}
 (I+T_{z,t})w_{z,t} =-T_{z,t} I.
 \end{equation}
Then functions the $q,\varphi$ defined in (\ref{1112A}) satisfy all the relations (\ref{dsiiB}) in the classical sense when $(z,t)\in\omega$.
  \item  Let us fix $z_0, t_0 \in \mathbb R^2\times \mathbb R^+$. Then for generic potential $q_0(z)$ in $W^{5,\infty}_{\rm{comp}}(\mathbb R^2)$, equation  (\ref{inteq})
  is uniquely solvable in $\mathcal H^{s}$ for all $(z,t)$ in some neighborhood of $(z_0,t_0)$. The neighborhood may depend on $q_0$.

\item
 Consider a set of initial data $aq_0(z)$ that depend on  $a \in (0,1]$. Then equation (\ref{inteq}) with $Q^0$ replaced by $a Q^0$ ($Q^0$ is fixed) is uniquely solvable for almost every $(z, t, a)\in\mathcal O \times \mathbb R^+ \times (0,1]$.  For each $(z,t)$, the unique solvability can be violated for at most finitely many values of $a=a_j(z,t),~z,t\in \mathcal O\times \mathbb R^+$.
   \end{itemize}
\end{theorem}

The proof consists of 3 parts. In section \ref{sec2}, we show (Lemma \ref{1112CL}) that the reconstruction procedure of the inverse scattering problem can be applied to an arbitrary regular and fast decreasing matrix-function $h$, which is not necessarily the scattering data $h_0$ of a compactly supported potential $Q^0$. This Lemma allows us to consider scattering data $h$ defined by (\ref{2506B}) for all $t\geq 0$, and construct operator $T_{z,t}$ and the solution $v_{z,t}$ of the equation (\ref{1603B}). As a result, one gets a potential $Q^t$ (which is not necessarily compactly supported) and a function $\psi$, which are defined in terms of $v_{z,t}$ for all $(z,t)\in\omega$ and are related by the Dirac equation $\frac{\partial \psi}{\partial \overline{z}} = Q^t \overline{\psi}$.

In section \ref{sec3}, we demonstrate that the solution $v_{z,t}$ of equation (\ref{1603B})
 satisfies the compatibility conditions. And, finally, in section \ref{compcond}, we repeat well-know arguments showing that the compatibility condition implies the validity of the first equation in the DSII problem  (\ref{dsiiB}), and complete the proof of Theorem \ref{mthm}.

\section{Inverse problems for general scattering data }\label{sec2}

Consider a $2\times 2$ matrix-function $h$ (non necessarily defined by (\ref{27dec2})) and a scalar function $w$ that satisfy the following condition.
\begin{eqnarray} \nonumber
\mbox{ {\bf Condition}.  1) $ h^o(\varsigma,\varsigma)\in L^\infty(\mathbb C \backslash D)\bigcap L^{2}(\mathbb C \backslash D)$  and   $h(\varsigma',\varsigma)\in L^\infty(\partial D) \times L^\infty(\partial D) $ } ,\\
2) ~ |w(\varsigma',\varsigma) | \leq C \left |Ln |\varsigma-\varsigma'| \right |, ~ \varsigma',\varsigma \in \partial D,\label{2602A}
\end{eqnarray}
where $h^o=\Pi^oh$ is the off-diagonal part of $h$. Sometimes we will need a stronger assumption on $h$:
\begin{equation}\label{strh}
|\varsigma|^3h^o(\varsigma,\varsigma)\in L^\infty(\mathbb C \backslash D)\bigcap L^{1}(\mathbb C \backslash D),  \quad  h(\varsigma',\varsigma)\in L^\infty(\partial D) \times L^\infty(\partial D).
\end{equation}

 Let $T_z$ be a slightly more general operator than those defined in (\ref{0712GAA}). Namely, consider the following operator $T_z$ acting in the space $\mathcal H^{{s}}, ~{s}>2$:
\begin{eqnarray}\nonumber
T_z \phi (k)=
\frac{1}{\pi} \int_{\mathbb C\backslash D}  e^{i(\overline{\varsigma}z+\overline{z}{\varsigma})/2} \overline{\phi}(\varsigma) \Pi^o h(\varsigma,\varsigma)  \frac{d\varsigma_R d\varsigma_I}{\varsigma-k}
\\ \label{glavOperator}
+\frac{1}{2\pi i} \int_{\partial D} \frac{d\varsigma}{\varsigma - k} \int_{\partial D}
[e^{i(\varsigma\overline{z}+\overline{\varsigma'}z)/2}\overline{\phi^-(\varsigma')} \Pi^o  + e^{i(\varsigma-{\varsigma'})\overline{z}/2}\phi^-(\varsigma')\Pi^d \bold C] \left [w(\varsigma',\varsigma)  h(\varsigma',\varsigma) d\overline{\varsigma'} \right ],
\end{eqnarray}
where $h$ and $w$ satisfy (\ref{2602A}),
$\phi^-$ is the boundary trace of $\phi$ from the interior of $D$,
$\bold C$ is the operator of complex conjugation. $\Pi^oM=M^o$ is the off-diagonal part of a matrix $M$ and $\Pi^dM=M^d$ is the diagonal part. Thus everywhere below $h$ is an arbitrary matrix satisfying (\ref{2602A}). The specific matrix $h=h_0$ defined in (\ref{27dec2}) via the initial data $q_0$ will appear only in Lemmas \ref{new1}, \ref{0312A} and at the very end of the last section of the paper.
\begin{lemma}\label{new1}
1) If (\ref{2602A}) holds, then operator $T_z$ is compact in $\mathcal H^{{s}}, ~{s}>2$, and
\begin{equation}\label{estt}
\|T_z\|_{\mathcal H^s}\leq C\|h\|_H, \quad \|h\|_H:=\|h^o(\varsigma,\varsigma)\|_{L^\infty(\mathbb C \backslash D)\bigcap L^{2}(\mathbb C \backslash D)}+\|h(\varsigma',\varsigma)\|_{L^\infty(\partial D) \times L^\infty(\partial D)}.
\end{equation}

2) If $h$ depends analytically on a complex parameter $\alpha\in A\subset \mathbb C$ or has $m\geq 0$ continuous derivatives with respect to a real parameter $\tau\in[0,T]$, where the derivatives in $\alpha, \tau$ are understood as derivatives of elements of the Banach space $H$, then operator $T_z$ is analytic in $\alpha$ and $m$ times differentiable in $\tau$, and the derivatives of the right hand side of (\ref{glavOperator}) can be moved inside the integrals.
\end{lemma}
{\bf Proof.} If $\mathcal E=\emptyset$ (there are no exceptional points), then $D$ is empty, $\mathcal H^s=L^s$, and operator $T_z$ can be simplfied significantly (in particular, the second line in (\ref{glavOperator}) can be dropped). In this case, the validity of the first statement of the lemma requires only $h^o\in L^2$, and the statement was proved by
Nachman \cite[Lemma 4.2]{nach} (the proof was reproduced in \cite[Lemma 5.3]{music}). In the general case, the proof can be found in \cite[Lemma 4.3]{lbcond}. In fact, the estimate (\ref{estt}) was not stated in \cite{lbcond} explicitly, but it can be easily extracted from the proof
of the compactness of $T_z$.

The validity of the second statement is obvious due to (\ref{estt}), and it is stated above solely for the convenience of references.

\qed

\begin{lemma}\label{new2}({\bf Sung \cite{sung1},\cite{sung2}})
If $q_0(z) \in W^{6,\infty}_{\rm{comp}}(\mathbb R^2)$, then (\ref{strh}) holds for the scattering matrix $h=h_0$ defined in (\ref{27dec2}).
\end{lemma}
{\bf Remark.} It was shown by Sung \cite{sung1},\cite{sung2} that condition
$q_0 \in W^{n,\infty}_{\rm{comp}}(\mathbb R^2)$ implies that $h_0(\varsigma,k)$ is a bounded continuous function on $\mathbb C \times (\mathbb C \backslash D)$ and
\begin{equation}\label{09}
\varsigma_1^{\beta_1} \varsigma_2^{\beta_2} \Pi^o h_0(\varsigma,\varsigma) \in C_0(\mathbb C \backslash D), ~ |\beta|\leq n, ~ \varsigma=(\varsigma_1,\varsigma_2),
\end{equation}
where $C_0(\mathbb C \backslash D)$ is the space of continuous matrix-functions with zero limit at infinity.
In fact, this inclusion is proved (see \cite[Lemma 2.16]{sung1}) when $\mathcal E=\emptyset$ and $D=\emptyset$, but the proof remains the same in the presence of exceptional points if $k\in \mathbb C \backslash D$.

\begin{lemma}\label{1112CL}
Let (\ref{strh}) hold. Then

1) $T_zI\in \mathcal H^{s}$ for each  ${s}>2$.

2) Let  $z$ be such  that the kernel of $I+T_z$ is trivial. Let $v=v(z,\cdot)$ be the unique solution of
\begin{equation}\label{gleq}
(I+T_z)v =I,
\end{equation}
such that $w=v-I$ satisfies $(I+T_z)w =T_zI\in \mathcal H^{s}$ and belongs to $H^{s}$.

Let
\begin{eqnarray}\nonumber
Q:=\frac{1}{2i}\Pi^o \mathcal Cv, \mbox{ where } \mathcal C v:=
 \frac{1}{\pi} \int_{\mathbb C\backslash D}  e^{i(\overline{\varsigma}z+\overline{z}{\varsigma})/2} \overline{v}(\varsigma) \Pi^oh(\varsigma,\varsigma)  {d\varsigma_R d\varsigma_I}    \\\label{0303C}
+\frac{1}{2\pi i} \int_{\partial D} {d\varsigma} \int_{\partial D}
[e^{i(\varsigma\overline{z}+\overline{\varsigma'}z)/2}\overline{v^-(\varsigma')} \Pi^o  + e^{i(\varsigma-{\varsigma'})
\overline{z}/2}v^-(\varsigma')\Pi^d \bold C]
\left [w(\varsigma',\varsigma) h(\varsigma',\varsigma) d\overline{\varsigma'} \right ] .\label{1112C}
\end{eqnarray}
Then
\begin{equation}\label{0303A}
\overline{\partial } \psi = Q\overline{\psi}, \quad \psi:=\Pi^d \overline{v}e^{i\overline{k}z/2} + e^{-i\overline z k/2} \Pi^o v,
\end{equation}
 and
\begin{equation}\label{0303B}
\partial \Phi = \dbar (\overline{Q} Q), \mbox{ where }  \Phi = \frac{1}{2i} \dbar \Pi^d \mathcal Cv.
\end{equation}

\end{lemma}
{\bf Remarks.} 1) Note that
\begin{equation}\label{2602B}
\cc v=\lim_{k \rightarrow \infty}k(v-I), \quad Q=\lim_{k \rightarrow \infty} \frac{-ik}{2}\Pi^o v.
\end{equation}
Relations (\ref{2602B}) can be easily checked if one replaces $v-I$ and $\Pi^ov$ from (\ref{gleq}) by $-T_zv$ and $-\Pi^oT_zv$, respectively.

 2) The statements and the proof remain the same if the space $L^1$ in (\ref{strh}) is replaced by $L^{\frac{2s}{2+s}}$.

{\bf Proof.}
The first statement is proved in \cite[Lemmas 4.2]{lbcond}. Lemma \ref{new1} above implies  that the operator $T_z$ is compact in $\mathcal H^s$, and therefore $I+T_z$ is Fredholm. Both properties (the validity of the first statement of the lemma and the compactness of $T_z$) are valid without the factor $|\varsigma|^3 $ at $h$ in (\ref{strh}). The presence of this factor is needed to guarantee that  $\partial^\alpha \dbar^\beta T_z, \alpha+\beta \leq 3,$ is compact in $\mathcal H^{s}$ and $\partial^\alpha \dbar^\beta  T_z I \in \mathcal H^{s} $ for each ${s}>2$.


The operator $T_z$ can be naturally split into two terms: $T_z=\m + \D$, where $\m$ involves integration over $\mathbb C \backslash D$ and $\D$ involves integration over $\partial D$. Thus, $T_z\phi=\m\overline{\phi}+\Pi^o\D\overline{\phi}+\Pi^d\D\phi$. The entries $M^{ij}, ~D^{ij}, ~i,j=1,2,$ of the matrix operators $\m$ and $\D$ are
$$
M^{12} \phi = \frac{1}{\pi} \int_{\mathbb C \backslash D} \frac{e^{i\Re (\overline{\alpha} z)} \phi(\alpha)h_{12}(\alpha,\alpha)}{ \alpha - k } d\alpha_\Re d \alpha_\Im ,
$$
\begin{equation}\label{0312B}
M^{21}\phi= \frac{1}{\pi} \int_{\mathbb C \backslash D} \frac{e^{i\Re (\overline{\alpha} z)} \phi(\alpha)h_{21}(\alpha,\alpha)}{ \alpha - k } d\alpha_\Re d \alpha_\Im,
\end{equation}

$$
D^{jj} \phi   = \frac{1}{2\pi i}\int_{\partial D}\frac{d\zeta}{\zeta -k}\int_{\partial D}
\overline{w(\varsigma,\varsigma')  h_{jj}(\varsigma',\varsigma) }e^{\frac{i}{2}(\varsigma-\varsigma')\overline z} \phi(\varsigma '){d\varsigma'}, \quad j=1,2,
$$
\begin{equation}\label{0312C}
D^{ij}\phi = \frac{1}{2\pi i} \int_{\partial D} \frac{d\zeta}{\zeta -k}\int_{\partial D} w(\varsigma,\varsigma')
  {h_{ij}(\varsigma',\varsigma)} e^{{\frac{i}{2}}(\varsigma \overline{z}+\overline{\varsigma'}z)}\phi(\varsigma ')d\overline{\varsigma'} ,\quad i\neq j.
\end{equation}

We rewrite the matrix equation (\ref{gleq}) as four equations for its components. The first row of the matrix equation is equivalent to
\begin{eqnarray}\label{2810A}
(I+D^{11})v_{11}+(M^{21}+D^{21})\overline{v_{12}}=1, \\ \label{2810B}
(I+D^{22})v_{12} +(M^{12}+D^{12}){\overline{v_{11}}}=0.
\end{eqnarray}
The second row of (\ref{gleq}) leads to similar equations for $v_{22}, v_{21}$.
This system of equations can be rewritten as the following four independent equations:
\begin{eqnarray}\label{eqA}
(I+D^{11})v_{11} - (M^{21}+D^{21})(I+\overline{D^{22}})^{-1}(\overline{M^{12}} + \overline{D^{12}})v_{11}=1, \\ \label{eqB}
(I+\overline{D^{22}})\overline{v_{12}} - (\overline{M^{12}}+\overline{D^{12}})(I+D^{11})^{-1} [(M^{21} +D^{21})\overline{v_{12}}-1]= 0,
\end{eqnarray}
\begin{eqnarray}\label{eqC}
(I+D^{22})v_{22} - (M^{12}+D^{12})(I+\overline{D^{11}})^{-1}(\overline{M^{21}} + \overline{D^{21}})v_{22}=1, \\ \label{eqD}
(I+\overline{D^{11}})\overline{v_{21}}- (\overline{M^{21}}+\overline{D^{21}})(I+D^{22})^{-1}[(M^{12} +D^{12})\overline{v_{21}}-1] =0,
\end{eqnarray}
if operators $(I+\overline{D^{ii}})$ are invertible. We will prove the lemma under this additional assumption on invertibility of $(I+\overline{D^{ii}})$, and we will get rid of this assumption at the very end of the proof.

By straightforward calculation, we get the following formulas for the derivatives of the operators $M^{ij}, ~D^{ij}$:
\begin{eqnarray}\label{0312F}
\frac{2}{i}\frac{\partial}{\partial z} (M^{ij}+D^{ij}) = (M^{ij}+D^{ij})\overline{X}, \quad i\neq j,\\
\frac{2}{i}\frac{\partial}{\partial z} (\overline{M^{ij}}+\overline{D^{ij}}) = C_{ij} -\overline{X}(\overline{M^{ij}}+\overline{D^{ij}}), \quad i\neq j, \\
\frac{\partial}{\partial z} D^{jj} =0, \\ \label{0312G}
\frac{2}{i}\frac{\partial}{\partial z} \overline{D^{jj}} = C_j - \overline{X}\overline{D^{jj}} + \overline{D^{jj}} \overline{X},
\end{eqnarray}
where $X$ is the operator of multiplication by the independent variable, $C_{ij},C_j$ are integral functionals that are closely related to the entries of the matrix $\mathcal C$ introduced in (\ref{1112C}):
\begin{eqnarray}
C_{ij}\phi =- \frac{1}{\pi} \int_{\mathbb C \backslash D} {e^{-i\Re (\overline{\alpha} z)}\overline{h_{ij}}(\alpha,\alpha)}\phi(\alpha)d\alpha_\Re d\alpha_\Im  \nonumber \\ \nonumber
+\frac{1}{2\pi i} \int_{\partial D}\int_{\partial D}  d\varsigma'\overline{d\varsigma w(\varsigma,\varsigma') h_{ij}(\varsigma',\varsigma)} e^{{\frac{-i}{2}}(\overline{\varsigma}z+\varsigma'\overline{z})}\phi(\varsigma '), \\
C_j\phi =\frac{1}{2\pi i} \int_{\partial D}\int_{\partial D}  d\overline{\varsigma'} {d\overline{\varsigma} {w(\varsigma,\varsigma')}h_{jj}(\varsigma',\varsigma)} e^{\frac{-i}{2}(\overline{\varsigma}-\overline{\varsigma'})z}\phi(\varsigma ').
\end{eqnarray}
The following relation is an immediate consequence of (\ref{0312G}):
\begin{equation}\label{inv}
\frac{2}{i}\frac{\partial}{\partial z} (I+\overline{D^{22}})^{-1} = -(I+\overline{D^{22}})^{-1}C_2(I+\overline{D^{22}})^{-1}  + (I+\overline{D^{22}})^{-1}\overline{X} - \overline{X}(I+\overline{D^{22}})^{-1}.
\end{equation}

Lets us differentiate (\ref{eqA}) using formulas (\ref{0312F})-(\ref{0312G}). We get
\begin{eqnarray}\nonumber
[(I+D^{11})- (M^{21}+D^{21})(I+\overline{D^{22}})^{-1}(\overline{M^{12}} +\overline{ D^{12}})]\frac{\partial v_{11}}{\partial z} = \\ \nonumber \frac{i}{2}(M^{21}+D^{21})(I+\overline{D^{22}})^{-1}[-C_2
(I+\overline{D^{22}})^{-1}(\overline{M^{12}} + \overline{D^{12}}) v_{11} + C_{12}v_{11}]= \\ \frac{i}{2}(M^{21}+D^{21})(I+\overline{D^{22}})^{-1}[C_2
 \overline{v_{12}} + C_{12}v_{11}]. \label{e30}
\end{eqnarray}
Note, that $c_0:=\frac{-i}{2}[C_2  \overline{v_{12}} + C_{12}v_{11}]$ does not depend on $k$. Let us replace $\frac{\partial v_{11}}{\partial z}$ in equation (\ref{e30}) by $v_{21}$ and omit the constant factor $c_0$ in its right-hand side. The resulting equation coincides with the equation obtained from (\ref{eqD}) by complex conjugation. Thus
\begin{equation}\label{2510E}
\frac{\partial v_{11}}{\partial z} = \overline{\mathcal Q_{12}} v_{21}, ~\mbox{ where} ~ \mathcal Q_{12} =\overline{c_0}= \frac{i}{2}\overline{[C_2  \overline{v_{12}} + C_{12}v_{11}]}.
\end{equation}
It is easy to see that this formula for $Q_{12}$ coincides with the one given in (\ref{1112C}).

The analysis for $v_{12}$ is similar.
We will differentiate (\ref{eqB}) by taking the derivatives of each term separately:
\begin{eqnarray}\nonumber
\frac{2}{i}\frac{\partial }{\partial z} (I+\overline{D^{22}})\overline{v_{12}} =  \frac{2}{i}(I+\overline{D^{22}}) \frac{\partial \overline{v_{12}}} {\partial z} + C_2\overline{v_{12}} - \overline{X}\overline{D^{22}} \overline{v_{12}} + \overline{D^{22}} \overline{X}\overline{v_{12}}= \\
(I+\overline{D^{22}}) \left (\frac{2}{i}\frac{\partial \overline{v_{12}}} {\partial z} + \overline{X} \overline{v_{12}} \right ) + C_2\overline{v_{12}} - \overline{X}(I+\overline{D^{22}}) \overline{v_{12}}.\label{2610A}
\end{eqnarray}
\begin{eqnarray}\nonumber
\frac{2}{i}\frac{\partial }{\partial z} \left [\overline{(M^{12}}+\overline{D^{12}})(I+D^{11})^{-1}\left [(M^{21} + D^{21})\overline{v_{12}} - 1 \right ] \right ]= \\ \nonumber
\frac{2}{i}(\overline{M^{12}}+\overline{D^{12}})(I+D^{11})^{-1}(M^{21} +D^{21}) \frac{\partial \overline{v_{12}}} {\partial z}-C_{12}v_{11} + \overline{X}(\overline{M^{12}}+\overline{D^{12}}) v_{11} \\
+ (\overline{M^{12}}+\overline{D^{12}})(I+D^{11})^{-1}(M^{21} + D^{21}) \overline{X} \overline{v_{12}}.\label{2610B}
\end{eqnarray}
Here we used the following consequence of (\ref{2810A}):
$$
v_{11}=-(I+D^{11})^{-1}\left [(M^{21} + D^{21})\overline{v_{12}} - 1 \right ].
$$
We multiply relations (\ref{2610A}), (\ref{2610B}) by $i/2$ and equate their right-hand sides (due to (\ref{eqB})). If we note that the last term in the right-hand side of  (\ref{2610A}) coincides with the term $\overline{X}(\overline{M^{12}}+\overline{D^{12}}) v_{11}$ in (\ref{2610B}) (due to (\ref{2810B})), then we arrive at the following equation
\[
[(I+\overline{D^{22}})+(\overline{M^{12}}+\overline{D^{12}})(I+D^{11})^{-1}(M^{21} + D^{21})]\left (\frac{\partial \overline{v_{12}}} {\partial z} + \frac{i}{2}\overline{X} \overline{v_{12}} \right )=c_0,
\]
where $c_0=\frac{-i}{2}[C_2  \overline{v_{12}} + C_{12}v_{11}]$ does not depend on $k$ (see (\ref{e30})). Let us replace $\frac{\partial \overline{v_{12}}} {\partial z} + \frac{i}{2}\overline{X} \overline{v_{12}} $ in the latter equation by $\overline{v_{22}}$. If we also divide its right-hand side by $c_0$, then the resulting equation will coincide with the equation obtained from (\ref{eqC}) by complex conjugation. Hence,
\begin{equation}\label{v12}
\left (\frac{\partial \overline{v_{12}}} {\partial z} + \frac{i}{2}\overline{X} \overline{v_{12}} \right )=\overline{\mathcal Q_{12}}\overline{v_{22}}, ~\mbox{ where} ~ \mathcal Q_{12} =\overline{c_0}= \frac{i}{2}\overline{[C_2  \overline{v_{12}} + C_{12}v_{11}]}.
\end{equation}
Similarly, we show that
\begin{equation}\label{2510E2}
\frac{\partial v_{22}}{\partial z} = \overline{\mathcal Q_{21}} v_{12}, ~\mbox{ where }  ~{\mathcal Q_{21}} = \lim_{k \rightarrow \infty} \frac{-ik}{2}v_{21},
\end{equation}
and finally
\begin{equation}\label{0612A}
\left (\frac{\partial {\overline{v_{21}}}} {\partial z} +\frac{i{\overline{X}}}{2} {\overline{v_{21}}} \right )={\overline{\mathcal Q_{21}}}{\overline{v_{11}}}.
\end{equation}

In order to complete the proof of (\ref{0303A}), it remains only to note that equations (\ref{2510E}), (\ref{v12}), (\ref{0612A}), and (\ref{2510E2}) for $v$ can be rewritten in the form $\overline{\partial } \psi = Q\overline{\psi}$ using the relation between $v$ and $\psi$ provided in the statement of the lemma.

Let us prove (\ref{0303B}).
From (\ref{2510E}) and (\ref{2510E2}), it follows that
\begin{equation}\label{2602C}
   \partial \Pi^d(v -1) = \Pi^d (\overline{Q} v).
\end{equation}
We multiply both sides of the above equation by $k$ and pass to the limit as $k \rightarrow \infty$ using (\ref{2602B}).  This implies
\begin{equation}\label{0103B}
\partial \Pi^d \mathcal C v =  2i \overline{Q} Q.
\end{equation}
Applying $\dbar$ and using $\partial \dbar = \dbar \partial$, we get
$$
\partial \left [\frac{1}{2i}\dbar\Pi^d \mathcal C v \right ] = \dbar  (\overline{Q}Q).
$$
This completes the proof of the lemma under the condition that operators $(I+\overline{D^{ii}})$ are invertible.

In order to prove Lemma \ref{1112CL} in the general case, we consider the scattering data $\gamma h$ instead of $h$, where $\gamma\in[0,1]$. Then operators $T_z$ and $\overline{D^{ii}}$ are analytic in $\gamma$ and vanish when $\gamma=0$. In order to prove the compactness of $T_z$ in \cite[Lemma 4.3]{lbcond}, we established the compactness of its components $\m$ and $\D$, i.e., $\overline{D^{ii}}$ were shown to be compact. Due to the analytic Fredholm theorem, operators $(I+\overline{D^{ii}})$ are invertible when $\gamma $ is close enough to one, $\gamma\neq 1$. Operator $I+T_z$ is invertible when $1-\gamma\ll 1$. Hence,   the relations (\ref{0303A}), (\ref{0303B})  hold when $0<1-\gamma\ll 1$. Since all the components of equalities (\ref{0303A}), (\ref{0303B}) are analytic in $\gamma$, these equalities holds for $\gamma=1$.

\qed


\section{Derivation of the compatibility condition }\label{sec3}
A symmetry of the potential $Q$ and scattering data $h$ will play an important role in this section. So, we start the section with two simple lemmas establishing the relation between those symmetries.
\begin{lemma}\label{0312H}
Let (\ref{strh}) hold and
\begin{equation}\label{hh}
h_{11}(\varsigma,k)=h_{22}(\varsigma,k), \quad h_{12}(\varsigma,k)=\pm h_{21}(\varsigma,k),
\end{equation}
for all the pairs $(\varsigma,k)=(k,k), ~k \in \mathbb C \backslash D$  or $(\varsigma,k) \in \partial D \times \partial D$.
Then for all $z \in \mathbb R^2$ such that the kernel of $(I+T_z)$ is trivial, the following symmetry relations are valid for the matrix $Q$ defined in  (\ref{0303C}) and solution $v$ of (\ref{gleq}):
\begin{equation}\label{qaq}
Q_{12}(z)=\pm Q_{21}(z),
\end{equation}
\begin{equation}\label{0103A}
v_{11}= v_{22}, \quad v_{12}=\pm v_{21}.
\end{equation}
\end{lemma}
(The converse statement is given in Lemma \ref{0312A}.)

{\bf Proof}.
It is enough to prove (\ref{0103A}).
Then (\ref{qaq}) follows from (\ref{2602B}).
Note that (\ref{0312B}) and (\ref{0312C}) imply that
\begin{equation}\label{qwe}
D^{11}=D^{22},\quad D^{12}=\pm D^{21},\quad M^{12}=\pm M^{21}.
\end{equation}

We rewrite equations (\ref{eqB}), (\ref{eqD}) in the form
$$
[(I+\overline{D^{22}})- (\overline{M}^{12}+\overline{D}^{12})(I+D^{11})^{-1}({M^{21}} + {D^{21}})]\overline{v_{12}} =-(\overline{M}^{12}+\overline{D}^{12})(I+D^{11})^{-1}1,
$$
$$
[(I+\overline{D^{11}})- (\overline{M}^{21}+\overline{D}^{21})(I+D^{22})^{-1}({M^{12}} + {D^{12}})]\overline{v_{21}} =-(\overline{M}^{21}+\overline{D}^{21})(I+D^{22})^{-1}1.
$$
From (\ref{qwe}) it follows that the coefficients for $v_{12},~v_{21}$ in these equations are equal to each other, and the right-hand sides differ by the factor $\pm 1$. This justifies the second relation in (\ref{0103A}). The first relation can be proved similarly using  (\ref{eqA}), (\ref{eqC}).

\qed

\begin{lemma}\label{0312A}
Let $Q_{12}=\pm Q_{21} \in L^p_{\rm{comp}}(\mathbb R^2), ~p>2,$  be compactly supported functions. Let $h=h_0$ be the scattering data defined in (\ref{27dec2}). Then
$$
h_{11}=h_{22}, \quad h_{12}=\pm h_{21}
$$
for $k \not \in \mathcal E$, where $\mathcal E$ is the set of exceptional points.
\end{lemma}
{\bf Proof.} Let
\begin{equation}\label{muSet5}
\mathcal L_k \varphi (z)= \frac{1}{\pi}\int_{\mathbb R^2} \varphi(w)\frac{e^{-i\Re(\overline{k}w)} dw_R dw_I}{z-w}.
\end{equation}
Then the Lippmann-Schwinger equation (\ref{19JanA}) after substitution (\ref{subs}) can be rewritten in the following form (see also \cite[(1.6)]{sung1}):
\begin{eqnarray*}
\mu_{11}=1 + \mathcal L_{k} [Q_{12}(z') \overline{\mu_{21}}(z',k)], \quad \mu_{21}= \mathcal L_{k} [Q_{21}(z') \overline{\mu_{11}}(z',k)],\\
\mu_{22}=1 + \mathcal L_{k} [Q_{21}(z') \overline{\mu_{12}}(z',k)], \quad \mu_{12}= \mathcal L_{k} [Q_{12}(z') \overline{\mu_{22}}(z',k)].
\end{eqnarray*}
Therefore,
\begin{eqnarray*}
\mu_{11}=1 + \mathcal L_{k} [Q_{12}(z') \overline{\mathcal L_{k}} [\overline{Q_{21}}(z') {\mu_{11}}(z',k)]],
\\
\mu_{21}= \mathcal L_{k} \left [ Q_{21}(z') \left ( 1 + \overline{\mathcal L_{k}} [\overline{Q_{12}}(z') {\mu_{21}}(z',k)] \right ) \right ],
\end{eqnarray*}
and
\begin{eqnarray*}
\mu_{22}=1 + \mathcal L_{k} [Q_{21}(z') \overline{\mathcal L_{k}} [\overline{Q_{12}}(z') {\mu_{22}}(z',k)]],
\\
\mu_{12}= \mathcal L_{k} \left [ Q_{12}(z') \left ( 1 + \overline{\mathcal L_{k}} [\overline{Q_{21}}(z') {\mu_{12}}(z',k)] \right ) \right ].
\end{eqnarray*}
So, we obtain
$$
\mu_{11}=\mu_{22}, \quad \mu_{12}=\pm \mu_{21}.
$$
Now Lemma \ref{0312A} follows from  (\ref{27dec2}).

\qed

Consider an arbitrary time-independent matrix $h$ that satisfies (\ref{strh}) and the symmetry condition (\ref{hh}). Using this matrix $h$, we define the following time dependent data
\begin{equation}\label{2506Bxxx}
h(\varsigma',\varsigma,t):=e^{-t(\varsigma^2-\overline{\varsigma'}^2)/2}\Pi^o h(\varsigma',\varsigma) +e^{-t(\overline{\varsigma}^2-\overline{\varsigma'}^2)/2}\Pi^d h(\varsigma',\varsigma), ~ \varsigma \in \mathbb C\backslash \mathcal E, ~ t \geq 0,
\end{equation}
and then apply Lemma \ref{1112CL} to construct, for each $t\geq 0$, the solution $v=v(z,k,t)$ of equation (\ref{gleq}) and the matrix $Q=Q^t(z)$ defined in terms of $\mathcal C$ in (\ref{1112C}). The symmetry condition (\ref{hh}) implies that the same symmetry condition holds for function (\ref{2506Bxxx}) for each $t>0$, and therefore, from Lemma \ref{0312H} it follows that
\begin{equation}\label{3010B}
Q^t_{12}(z)=\pm Q^t_{21}(z) =:q(z,t),
\end{equation}
and that  (\ref{0103A}) holds. We denote by $\varphi$ the diagonal entries of the matrix $\Phi=\frac{1}{2i} \dbar \Pi^d \mathcal C v$ defined in (\ref{0303B}). These entries are equal due to (\ref{0103A}) and (\ref{2602B}). The proof of the latter statement requires changing the order of the operator $\dbar$ and the limit in (\ref{2602B}). One can give an elementary alternative proof using the definition (\ref{1112C}) of matrix $\mathcal C$ and the symmetries of $h$ and $v$. Thus (\ref{0303B}) takes the form
\begin{equation}\label{0103D}
\partial \varphi=\dbar |q|^2.
\end{equation}

We will show that $v$ satisfies the compatibility conditions (see \cite[(1.26)]{sung1}, or Section \ref{compcond}):
\begin{equation}\label{2410A}
\frac{\partial v}{\partial t} +2\left[\frac{\partial^2 v}{\partial \overline{z}^2} + \frac{\partial^2 v}{\partial {z}^2} \right ] - 2ik \frac{\partial v}{\partial \overline{z}} + A(\Pi^d v) + \overline{A}(\Pi^o v) =0,
\end{equation}
where the entries of the matrix $A=A(z,t)$ are
\begin{eqnarray}\label{2510B}
A_{12}=\pm A_{21} = -4\overline{ \partial } q, \\ \label{2510C}
A_{11}=A_{22}= \mp 4\varphi.
\end{eqnarray}

\begin{lemma}\label{1112B}
Let matrix $h$ have properties (\ref{strh}) and (\ref{hh}), and let $T=T_{z,t}$ be the operator (\ref{glavOperator}) with $h$ defined in (\ref{2506Bxxx}).  Then (\ref{2410A}) holds in the classical sense at each $(z,t)$ for which the kernel of $I+T_{z,t}$ is trivial.
\end{lemma}
{\bf Remark.} We will prove later that the vector $(q,\varphi)$, where $q$ is defined by (\ref{3010B}) and $\varphi$ is defined immediately after (\ref{3010B}), satisfies the DSII equation.

{\bf Proof.}
As earlier, we split operator (\ref{glavOperator}) as follows: $T=\m + \D^o+\D^d$, where $\m$ involves integration over $\mathbb C \backslash D$, and $\D=\D^o+\D^d$ involves integration over $\partial D$.  Similarly, we split $\mathcal Cv$ in three natural terms
\begin{eqnarray}\nonumber
\cc v=(\cc^1 + \cc^o + \cc^d )v=
\frac{1}{\pi} \int_{\mathbb C\backslash D}  e^{i(\overline{\varsigma}z+\overline{z}{\varsigma})/2} \overline{v}(\varsigma) \Pi^oh(\varsigma,\varsigma,t)  {d\varsigma_R d\varsigma_I} \\
\label{xxy}
+\frac{1}{2\pi i} \int_{\partial D} {d\varsigma} \int_{\partial D}
[e^{i/2(\varsigma\overline{z}+\overline{\varsigma'}z)}\overline{v^-(\varsigma')} \Pi^o  + e^{i/2(\varsigma-{\varsigma'})\overline{z}}v^-(\varsigma')\Pi^d \bold C] \left [{\rm Ln}\frac{\overline{\varsigma'}-\overline{\varsigma} }{\overline{\varsigma'}-\overline{k_0} }h(\varsigma',\varsigma,t) d\varsigma' \right ].
\end{eqnarray}

{\it Warning.} We are using the standard notation $T\phi$ for the action of an operator $T$ on a matrix $\phi$. However, one must keep in mind that the products of matrices in the integrands in (\ref{glavOperator}) are taken in an unusual order with the factor $\phi$ being on the left.

 We will use the fact that
\begin{equation}\label{0501A}
\Pi^d [(I+T)^{-1}(iI)] = i[\Pi^d (I+T)^{-1}I]  \quad \mbox{ and }\quad  \Pi^o[ (I+T)^{-1}(iI)]= -i\Pi^o [(I+T)^{-1}I].
\end{equation}
One can justify (\ref{0501A}) by checking that the elements of the matrix $i\Pi^dv-i\Pi^ov$ satisfy (\ref{eqA})-(\ref{eqD}) if the latter equations are multiplied by $i$.

Since $(I+T)v=I$ and matrix $A$ does not depend on $k$, it follows from (\ref{0501A}) and the warning above that
$$
(I+T)(A(\Pi^d v) + \overline{A}(\Pi^o v))=A .
$$
Since the kernel of the operator $I+T$ is trivial, one can check that $I+T$ applied to the left-hand side of (\ref{2410A}) is zero, instead of proving (\ref{2410A}). Thus, in order to prove Lemma \ref{1112B}, it is enough to show that
\begin{equation}\label{yyy}
(I+T)(v_t +2Lv - 2ik v_{\overline z})+A=0, \quad  Lv:=v_{\overline{z} \overline{z}}+v_{zz}=\frac{1}{2}(v_{xx}-v_{yy}).
\end{equation}

Applying operators $\frac{\partial}{\partial t} $ and $L$ to the relation $(I+T)v=I$, we obtain that
\[
(I+T)v_t=-T_tv, \quad  (I+T)Lv=-(LT)v-(T_x v_x-T_y v_y).
\]
This allows us to rewrite (\ref{yyy}) in the form
\begin{equation}\label{2510A}
-(T_t+2T_{\overline{z}\overline{z}}+2T_{zz})v -2(T_x v_x-T_y v_y) - (I+T)(2ikv_{\overline z})+A=0.
\end{equation}

We split the proof of (\ref{2510A}) into several steps. Note that by a straightforward calculation (using (\ref{glavOperator}) and formula (\ref{1112C}) for $\mathcal C$), one can verify that
 \begin{equation}\label{AA}
(T_t+2T_{\overline{z}\overline{z}}+2T_{zz}-2ikT_{\overline{z}})v= 2i\cc_{\overline{z}} v.
\end{equation}
Recall that $T$ is not linear with respect to multiplication by $i$, so the term $(I+T)(2ikv_z)$ in (\ref{2510A}) needs an accurate treatment.
We will show that
\begin{equation}\label{2410C}
 -2(T_x v_x-T_y v_y) -(I+T)(2ikv_{\overline z})=2ikT_{\overline{z}}v -2i(\cc^1 v_z + \cc^o v_z +\cc^d v_{\overline{z}}).
 \end{equation}
It will be also shown that
\begin{equation}\label{2410D}
A=2i\overline{\partial} (\cc v).
\end{equation}
Since both functions $v$ and $\overline{v}$ are present in integrands in the right-hand side of (\ref{xxy}), formula (\ref{2410D}) is equivalent to $A=2i (\cc_{\overline{z} } v+\cc^1 v_z + \cc^o v_z +\cc^d v_{\overline{z}})$. Hence, (\ref{AA})-(\ref{2410D}) justify (\ref{2510A}), and therefore Lemma \ref{1112B} will be proved as soon as (\ref{2410C}), (\ref{2410D}) are established.

The equality of the diagonal terms in (\ref{2410D}) follows from (\ref{2510C}) and the definition of $\varphi$. In order to verify the validity of (\ref{2410D}) for the non-diagonal elements, we substitute the left-hand side of (\ref{3010B}) for $q$ in (\ref{2510B}) and use the relation $2i \Pi^o(\cc v)=-4\Pi^oQ$, which follows from (\ref{2602B}). Thus  (\ref{2410D}) is proved, and it remains only to prove (\ref{2410C}).

We apply operator $\overline{\partial}$ to (\ref{gleq}) and then multiply the resulting equation by $-2ik$:
\begin{equation}\label{2410C2}
-2ik v_{\overline{z}} - 2ik ((\m+\D^o)v_z + \D^d v_{\overline{z}}) = 2ik (\m_{\overline{z}}v +\D^o_{\overline{z}}v + \D^d_{\overline{z}}v).
\end{equation}
Now we will rearrange each term in (\ref{2410C2}) separately, keeping in mind that $k$ is an independent variable, and multiplication by $k$ does not commute with the operators above. The rearrangement will involve an additional term $T_xv_x-T_yv_y$.

{\bf Operator $\m$}.
We will use the fact that $i\m(\cdot)=\m(-i\cdot)$. We have
\begin{eqnarray}\label{0501B}
\m(2i\overline{k} v_z) = \m((i\Re k + \Im k)(v_x -  i v_y)) = \m(i(\Re k v_x - \Im k v_y) + (\Im k v_x + \Re k v_y)), \\ \label{0501C}
\m(2i{k} v_{\overline z}) = \m(i(\Re k v_x - \Im k v_y) - (\Im k v_x + \Re k v_y)), \\ \label{0501D}
-\m_x v_x+ \m_y v_y = \m (i(\Re k v_x - \Im k v_y)).
\end{eqnarray}
From (\ref{0501B})-(\ref{0501D}) it follows that
\begin{equation}\label{2810E1}
2[-\m_x v_x + \m_y v_y]- \m(2i{k} v_{\overline z}) = \m(2i\overline{k} v_z) = -2ik \m v_z - 2i \cc^1 v_z.
\end{equation}

{\bf Operator $\D^o$}. Similar formulas hold for $\D^o$. We will use the fact that $i\D^o(\cdot)=\D^o(-i\cdot)$. The exponent in the term of (\ref{glavOperator}) that corresponds to $\D^o$ can be rewritten as follows:
$$
\frac{i}{2} (\varsigma \overline{z}+ \overline{\varsigma'} z) = ix \frac{\varsigma+\overline{\varsigma'}}{2} + y \frac{\varsigma-\overline{\varsigma'}}{2}.
$$
One can check that
$$
2[- \D^o_x v_x+ \D^o_y v_y] = - ik \D^o v_x  + \D^o (i k  v_x) - \D^o (k   v_y) + k \D^o v_y + [-i\cc^o v_x +  \cc^o v_y].
$$

Note that
$$
\D^o(2i{k} v_{\overline z}) = \D^o(i{k} v_x - k v_y).
$$
From the last two equalities it follows that
\begin{eqnarray} \nonumber
2[- \D^o_x v_x +  \D^o_y v_y]-  \D^o (2i{k} v_{\overline z}) = [-i\cc^o v_x +  \cc^o v_y]- ik  \D^o v_x +k\D^o v_y  \\ \nonumber = -i[\cc^o v_x  +i\cc^o v_y)]
-ik(\D^o  v_x +i\D^o v_y)= -i\cc^o(v_x -iv_y) -ik\D^o (v_x -iv_y)\\= -2i \mathcal C^o v_{z} - 2ik \D^o v_z. \label{2810E2}
\end{eqnarray}

{\bf Operator $\D^d$.} Let us rearrange the exponent in the term of (\ref{glavOperator}) that corresponds to $\D^d$:
$$
\frac{i}{2} (\varsigma - {\varsigma'}) \overline z = i x\frac{\varsigma-{\varsigma'}}{2} + y \frac{\varsigma-{\varsigma'}}{2}.
$$
Then
$$
2[- \D^d_x v_x + \D^d_y v_y] = - ik \D^d v_x  +\D^d ( i k v_x) - \D^d (kv_y)  + k\D^d v_y - [i\cc^d v_x - \cc^d v_y].
$$

Note also that
$$
\D^d  (2i{k} v_{\overline z})= \D^d(i{k} v_x - k v_y).
$$
From the last two equalities we get
\begin{eqnarray}\nonumber
2[-\D^d_x v_x+ \D^d_y v_y]-  \D^d(2i{k} v_{\overline z}) = [-i\cc^d v_x + \cc^d v_y]- ik \D^d v_x +k\D^d v_y \\ \nonumber =
-i(\cc^d v_x + i\cc^d v_y) -ik(\D^d v_x  +i\D^d v_y)= -i\cc^d(v_x +iv_y) -ik\D^d(v_x +iv_y))\\= -2i \cc^d v_{\overline z}  -
2ik \D^dv_{\overline z}. \label{2810E3}
\end{eqnarray}

Now (\ref{2410C}) follows from (\ref{2410C2}) combined with (\ref{2810E1}),  (\ref{2810E2}), and (\ref{2810E3}).

\qed

\section{Proof of Theorem \ref{mthm}}\label{compcond}

We start with a theorem showing that the compatibility condition for the solution $v$ of (\ref{1603B}) implies that the functions $q,\varphi$ defined by $v$ satisfy the first two relations of the DSII system (\ref{dsiiB}):
\begin{equation}\label{2710D}
q_t=2(\overline{\partial}^2- \partial^2) q \pm 4q (\overline{\varphi}-\varphi), \quad \partial \varphi = \dbar |q|^2.
\end{equation}
\begin{theorem}\label{0712B} Let $h$ be an arbitrary scattering data that depends on $t\geq 0$ and, for each $t$, satisfies (\ref{strh}) and the symmetry relations (\ref{hh}). Let the kernel of operator $I+T_{z,t}$ be trivial in a neighborhood $\omega$ of a point $(z_0,t_0), ~ z_0 \in \mathbb C, ~t_0 \geq 0$. If the solution $v=v_{z,t}$ of equation (\ref{1603B}) satisfies the compatibility condition (\ref{2410A}) in $\omega$, then the function $q$ defined in (\ref{3010B}) and the diagonal elements $\varphi$ of the matrix $\Phi$  satisfy (\ref{2710D}) in $\omega$.
\end{theorem}
{\bf Proof.} One needs to prove only the first relation in (\ref{2710D}), since the second one was proved in (\ref{0103D}).

  We take the complex conjugate of (\ref{0612A}), differentiate it in $t$, and replace the derivatives of $v_{ij}$ using (\ref{2410A}). We get
\begin{eqnarray}\nonumber
(\overline \partial -\frac{ik}{2})\left ( -2 ({\partial}^2+ \overline{\partial^2})v_{21}+2ik \overline{\partial} v_{21} \pm 4 \overline{\partial} q   v_{11} \pm 4 \overline{\varphi}   v_{21}
\right ) \\ = \pm q_t v_{11}  \pm q \left ( -2 ({\partial}^2+ \overline{\partial^2})v_{11}+2ik \overline{\partial} v_{11} \pm 4 \varphi v_{11} + 4 \partial\overline{q} v_{21}
\right ), \quad k\in \mathcal C\setminus D.\nonumber
\end{eqnarray}
Hence, the theorem will be proved if we show that
\begin{eqnarray}\nonumber
(\overline \partial -\frac{ik}{2})\left ( -2 ({\partial}^2+ \overline{\partial^2})v_{21}+2ik \overline{\partial} v_{21} \pm 4 \overline{\partial} q  v_{11} \pm 4 \overline{\varphi}   v_{21}
\right ) \\ \nonumber  =\pm\left( 2(\overline{\partial}^2- \partial^2) q \pm  4q (\overline{\varphi}   - \varphi) \right ) v_{11} \\ \pm q \left ( -2 ({\partial}^2+ \overline{\partial^2})v_{11}+2ik \overline{\partial} v_{11} \pm 4 \varphi v_{11} + 4 \partial\overline{q} v_{21}
\right ), \quad k\in \mathcal C\setminus D.\label{0712C}
\end{eqnarray}
Indeed, the difference between the last two equations is equal to the difference between the left and right hand sides in the first of equations (\ref{2710D}) multiplied by $\pm v_{11}$. Thus (\ref{0712C}) implies that either (\ref{2710D}) holds or  $v_{11}=0$. We note that the terms in (\ref{2710D}) do not depend on $k$, and $v_{11}\to 1$ as $|k|\to \infty$. Thus the validity of (\ref{0712C}) implies (\ref{2710D}).

We apply the operator $\overline \partial -\frac{ik}{2}$ to all the terms in the left-hand side of (\ref{0712C}) and then open all the brackets except the ones in the expression $(\overline \partial -\frac{ik}{2})$. Then the left-hand side will have five terms, which will be denoted by $A^i$, and the second and third lines in (\ref{0712C}) will have four terms $B^i, 1\leq i\leq 4,$ and five terms $C^i, 1\leq i\leq 5$, respectively.
Thus we need to show that
\begin{equation}\label{222}
\sum_{i=1}^5 A^i = \sum_{i=1}^4 B^i  + \sum_{i=1}^5 C^i.
\end{equation}
We apply the complex conjugation to equation (\ref{0612A}). Using the resulting equation, we obtain
\begin{equation}\label{3010F}\nonumber
A^1= -2\partial^2(\pm q v_{11}) = \mp 2(\partial^2{q})v_{11} \mp 4(\partial {q}) \partial v_{11} \mp 2{q} (\partial^2 v_{11}) =:\sum_{i=1}^3 A^1_i.
\end{equation}
\begin{equation}\label{3010G} \nonumber
A^2=-2\overline{\partial}^2 (\pm q v_{11}) = \mp 2(\overline{\partial}^2{q})v_{11} \mp 4(\overline{\partial } {q}) \overline{\partial }v_{11} \mp2{q} (\overline \partial^2 v_{11})=:\sum_{i=1}^3 A^2_i.
\end{equation}
$$
A^3 =2ik \overline{\partial} (\pm{q}v_{11})=\pm 2ik (\overline{\partial} {q})v_{11} \pm  2ik {q} (\overline{\partial} v_{11}) =: A^3_1+A^3_2.
$$
$$
 A^4 = \pm4(\overline \partial -\frac{ik}{2})  (\overline{\partial} q  v_{11} )= \pm4 ((\overline \partial -\frac{ik}{2})  \overline{\partial} q  )v_{11} \pm4 (\overline{\partial} q  )(\overline \partial v_{11}) = :A^4_1+A^4_2.
 $$
 $$
 A^4_1  = \pm4 ( \overline{\partial}^2 q  )v_{11} \mp 2ik  \overline{\partial} q  v_{11} = :A^4_{1,1}+ A^4_{1,2}.
 $$
$$
 A^5 = \pm4(\overline \partial -\frac{ik}{2}) ( \overline{\varphi}  v_{21}) = \pm 4(\overline{\partial}  \overline{\varphi} ) v_{21}
 +4 \overline{\varphi} q v_{11}= :A^5_1+A^5_2,
 $$
 where, due to (\ref{0103D}),
\begin{equation}\label{3010I}\nonumber
A^5_1 = \pm 4v_{21} {\partial} |q|^2 = :\pm 4v_{21}(q {\partial} \overline{q} + \overline{q} {\partial}q)=A^5_{1,1}+A^5_{1,2}.
\end{equation}

From (\ref{2510E}) it follows that $A^1_2+A^5_{1,2}=0$. Other relations below can be easily checked. Together, they prove (\ref{222}).
\begin{eqnarray}\label{0712E}\nonumber
A^1_1 = B^2, \quad A^1_3+A^2_3=C^1+C^2, \quad A^2_1+A^4_{1,1} = B^1, \quad A^2_2 + A^4_2=0,
\\ \nonumber
A^3_1+A^4_{1,2}=0, \quad A^3_2=C^3, \quad A^5_{1,1}=C^5,\quad A^5_2=B^3, \quad B^4+ C^4=0.
\end{eqnarray}

\qed


{\bf Proof of Theorem \ref{mthm}}.
The compactness of $T_{z,t}$ and its derivatives stated in the first item of the theorem follows from Lemmas \ref{new1}, \ref{new2}. Note that condition (\ref{2602A}) implies the compactness of $T_{z,t}$, and (\ref{strh}) is needed for the compactness of its derivatives. The inclusion $T_{z,t}I\in\mathcal H^s, ~s>2,$ is
due to the Hardy-Littlewood-Sobolev inequality, more details can be found in Lemma 4.2 of \cite{lbcond}.

From Lemma \ref{1112B} and Theorem \ref{0712B} it follows that the functions $q,\varphi$ defined in (\ref{1112A}) satisfy the first two relations of (\ref{dsiiB}) when $(z,t)\in\omega$. Let us justify the validity of the initial condition (\ref{dsiiB}). Note that $q(x,0)$ is given by (\ref{1112A}) and coincides with the elements $q_{12}$ of the matrix $Q$ in (\ref{gleq}) if the operator $T_z$ in (\ref{gleq}) is the same as in (\ref{0712GAA}). If $T_z$ is given by (\ref{0712GAA}), then $q_{12}=q_0$ due to Theorem 2.1 from \cite{lbcond}.

The third item of the theorem, i.e., that equation (\ref{1112A}) is solvable in a neighborhood of each point $(x_0,t_0)$ for generic potentials, is proved in Theorem 2.1 of \cite{lbcond} when $t=t_0$ is fixed. This proof remains valid when $t$ is close to $t_0$. The validity of the last statement of the theorem is justified in Remark 2 after Theorem 2.1. of \cite{lbcond}.

\qed

{\bf Acknowledgments.} This paper would not have been written without productive consultations with R.G. Novikov. The authors are also  grateful to Paula Cerejeira, Uwe Kahler, Anna Kazeykina and Jean-Claude Saut for discussions and kind support.

\bibliographystyle{amsalpha}

\end{document}